\documentclass[fleqn,usenatbib]{mnras}

\usepackage{newtxtext,newtxmath}
\usepackage{breqn}

\usepackage[T1]{fontenc}
\usepackage{ae,aecompl}
\usepackage[dvipsnames]{xcolor}
\usepackage[toc,page]{appendix}
\usepackage{pdflscape}
\usepackage{longtable}
\usepackage{booktabs}

\usepackage{graphicx} 
\usepackage{amsmath} 
\usepackage{rotating}

\title[Optical and X-ray Detection of a Thermonuclear Burst]{Simultaneous Optical and X-ray Detection of a Thermonuclear Burst in the 2024 Outburst of EXO 0748--676}

\author[A. H. Knight et al.]{
Amy H. Knight,$^{1}$\thanks{E-mail: amy.h.knight@durham.ac.uk}
Lauren Rhodes,$^{2}$
Douglas J. K. Buisson, $^{3}$
James H. Matthews,$^{2}$
Noel Castro Segura,$^{4}$
\and \ Adam Ingram,$^{5}$ 
Matthew Middleton,$^{6}$
and Timothy P. Roberts $^{1}$ \\
\\
$^{1}$ Centre for Extragalactic Astronomy, Department of Physics, Durham University, South Road, Durham DH1 3LE, UK \\
$^{2}$Department of Physics, Astrophysics, University of Oxford, Denys Wilkinson Building, Keble Road, Oxford, OX1 3RH, UK\\
$^{3}$Independent Researcher\\
$^{4}$Department of Physics, Gibbet Hill Road, University of Warwick, Coventry, CV4 7AL, United Kingdom\\
$^{5}$School of Mathematics, Statistics, and Physics, Newcastle University, Newcastle upon Tyne, NE1 7RU, UK\\
$^{6}$School of Physics and Astronomy, University of Southampton, Highfield, Southampton, SO17 1BJ, UK\\}

\date{Accepted XXX. Received YYY; in original form ZZZ}

\pubyear{2024}

\begin{document}
\label{firstpage}
\pagerange{\pageref{firstpage}--\pageref{lastpage}}
\maketitle

\begin{abstract}
The neutron star low-mass X-ray binary, EXO 0748--676, recently returned to outburst after a $\sim 16$ year-long quiescence. Since its return, there has been a global effort to capture the previously unseen rise of the source and to understand its somewhat early return to outburst, as it is typical for a source to spend longer in quiescence than in outburst. Here, we report on the simultaneous optical and X-ray detection of a type I X-ray burst, captured by \textit{XMM-Newton} during a DDT observation on 30th June 2024. The data show $3$ X-ray eclipses consistent with the known ephemeris and one type I X-ray burst at  $60492.309$ MJD. The X-ray burst is reprocessed into the optical band and captured by \textit{XMM-Newton's} Optical Monitor during a $4399$ s exposure with the B filter in image $+$ fast mode. We determine that the optical peak lags the X-ray peak by $4.46 \pm 1.71$s. The optical and X-ray rise times are similar, but the optical decay timescale is shorter than the X-ray decay timescale. The reprocessing site is likely within a few light seconds of the X-ray emitting region, so the companion star, accretion disc and ablated material are all plausible.
\end{abstract}



\vspace{-0.5cm}
\section{Introduction}
The neutron star (NS) low mass X-ray binary (LMXB) EXO 0748--676 was initially discovered as an X-ray transient in 1985 by the European X-ray Observatory Satellite (\textit{EXOSAT}; \citealt{Parmar1986}). Its optical counterpart, UY Vol, was discovered shortly after the initial X-ray detection \citep{Wade1985}. EXO 0748--676 exhibited an unusually long X-ray outburst lasting $\sim 24$ years before entering X-ray quiescence in late 2008 \citep{Hynes2009, Degenaar2011}. During its first outburst (1985 - 2008), EXO 0748--676 underwent regular monitoring at X-ray wavelengths by \textit{RXTE} \citep{Wolff2009, Knight2023} and \textit{XMM-Newton}, uncovering X-ray eclipses lasting $t_e \approx 500$ s that recur on the orbital period of $P=3.824$ hrs \citep{Parmar1986, Parmar1991, Wolff2009, Knight2022a, Knight2023} and type I X-ray bursts which established that the accretor in EXO 0748--676 is a NS \citep{Gottwald1986, Wolff2005, Boirin2007, Galloway2008}. Further analysis of the X-ray bursts determined that the system resides at a distance of $7.1 \pm 0.14$ kpc \citep{Galloway2008b}. 

The X-ray eclipses arise when the $\sim 0.4 M_{\odot}$ companion star obscures our view of the $\sim 2 M_{\odot}$ NS \citep{Ozel2006, Knight2022a}, thus confirming that the system is highly inclined ($\sim 76^{\circ}$; \citealt{Knight2022a}) and justifies the presence of significant dipping activity, which is due to the occultation of the X-ray emitting region by an accretion disc wind or similar, optically thick structure. Corresponding optical eclipses occur in phase with the X-ray eclipses, although they are longer in duration \citep{Crampton1986}. Detailed analysis of the X-ray eclipses observed by \textit{XMM-Newton} and \textit{RXTE} revealed that the M-dwarf companion star is undergoing ablation by X-ray irradiation and/or a pulsar wind from the NS. The excess ablated material introduces energy dependence and asymmetry into the observed X-ray eclipse profiles, similar to radio eclipses observed in spider pulsars \citep{Knight2022a, Knight2023}. Subsequently, the discovery of the reversal of the X-ray eclipse asymmetry, which is due to the movement of ablated material around the system and a phenomenon only previously observed in the black widow pulsar PRS J2051$-$0827 \citep{Polzin2019a}, solidified the sub-classification of EXO 0748--676 as a false widow, \citep{Knight2023} and implied that it would eventually evolve into a spider pulsar. Currently, there are no published detections of X-ray or radio pulsations at or around the expected NS spin frequency of $552$ Hz \citep{Galloway2010}. However, such pulsations are expected in false widows as they are considered an intermediate evolutionary state between NS LMXBs and spider pulsars \citep{Knight2023}.

EXO 0748--676 indicated its return to outburst on 10th June 2024, after $\sim 16$ years in quiescence \citep{Baglio2024, Rhodes2024}, with a type I X-ray burst \citep{Kuulkers2024} that triggered Swift-BAT with a peak count rate of $\sim 500$ cts/sec in the 15-350 keV band (GCN \#36653). As X-ray binaries typically spend more time in quiescence than outburst, EXO 0748--676's return was somewhat unexpected, but reminiscent of the short quiescences seen in the VY Scl cataclysmic variable sub-class \citep{Leach1999}. Following this initial burst, EXO 0748--676 was quickly observed by \textit{NICER} and \textit{NuSTAR} which captured several more type I X-ray bursts \citep{Kuulkers2024, Aoyama2024, Mihara2024, Bhattacharya2024Atel, Bhattachrarya2024Paper,Kashyap2024} and X-ray eclipses \citep{Buisson2024}, the latter of which confirmed the orbital ephemeris. Approximately two months prior, the ATLAS survey captured EXO 0748--676 rising in brightness \citep{Rhodes2024} and displayed significant optical variability, with average magnitudes of 18.1 in the 420-640nm band, and 17.7 between 560-820nm. On 30th June 2024, \textit{XMM-Newton} conducted a simultaneous optical and X-ray observation of EXO 0748--676 through director's discretionary time (DDT ObsID: 0935191701, PI: Dr. Norbert Schartel), which captured three eclipses and one type I X-ray burst. Here, we report the first simultaneous optical and X-ray detection of a type I X-ray burst captured by \textit{XMM-Newton} during the new outburst of EXO 0748--676.

Type I X-ray bursts result from thermonuclear fusion in the layers of accreted material that build up on the surface of NSs in X-ray binaries (XRBs). Upon achieving sufficient temperature and density, the surface material ignites, resulting in observable X-ray bursts that present as prompt increases in the observed X-ray flux and typically peak at levels significantly brighter than that of the persistent X-ray emission \citep{Lewin1993, Strohmayer2006, Galloway2008, Galloway2021}. X-ray bursts are highly energetic, thus their emission heats the outer accretion disk and the companion star, which subsequently, is reprocessed into optical emission (see e.g. \citealt{Hackwell1979, Lawerence1983, Matsuoka1984, Hynes2006, Paul2012}). The  optical emission appears after a short delay due to the reprocessing and light travel times. Similar lags (also known as echoes) are present in the absence of bursts \citep{Obrien2002,Vincentelli2023}. The optical emission presents similarly to the X-ray emission, brightening by $ 1 - 2$ magnitudes during a burst and often exhibits the same fast rise -- exponential decay (FRED) that their X-ray counterparts show \citep{Degenaar2018, Paul2012}. However, modern levels of sensitivity show type I burst profiles to be more complex. For example, they can display two-step rises \citep{Albayati2021}, so FRED models are not always suitable.
 
Simultaneous X-ray and optical bursts have only been detected in a small number of systems including Ser X-1 \citep{Hackwell1979}, 4U MXB 1636-53 \citep{Pedersen1982,Lawerence1983, Matsuoka1984} and EXO 0748-676 \citep{Hynes2006, Paul2012}. Uniquely, the eclipsing nature of EXO 0748--676 means that simultaneous observations at X-ray and optical wavelengths can probe the system's geometry and reprocessing mechanisms as a function of the orbital phase. \citet{Paul2012} previously explored this using a subset \textit{XMM-Newton} observations of EXO 0748--676 from 2001, noting that if the irradiated face of the companion star is reprocessing the X-ray bursts then one would expect strong orbital phase dependence in the reprocessing parameters. However, they do not find significant orbital phase dependence. While this argues against the irradiated face of the companion star being the reprocessing site, significant uncertainty remains, and other sites, such as the accretion disc or the ablated outflow are plausible. Further observations and analysis are required to distinguish between reprocessing sites.

\section{XMM-Newton Observations and Reduction}
\textit{XMM-Newton} observed EXO 0748--676 on 2024-06-30 starting at 14:30:35 UTC and ending on 2024-07-01 at 12:42:15 (ObsID: 0935191701, PI: Dr. Norbert Schartel), utilising all on-board instruments (see \citealt{Bhattachrarya2024Paper} for discussion of the high resolution spectroscopy from RGS). Details of the exposures considered here are provided in Table \ref{tb:XMMObs}. 

\begin{table}
\begin{center}
\begin{tabular}{ c c c c } 
Instrument & Mode & Filter & Duration(s) [s] \\ 
\hline
\vspace*{0.1cm}
EPIC MOS1 & Small Window & Thick & 54223\\ 
\hline
\vspace*{0.1cm}
EPIC MOS2 & Full Frame & Thick & 54210 \\ 
\hline
\vspace*{0.1cm}
EPIC PN & Timing & Thick & 49309\\ 
\hline
\vspace*{0.1cm}
OM & Image + Fast & V & 2120, 4399 \\
& & & 4401, 4399 \\ 
\hline
\vspace*{0.1cm}
OM & Image + Fast & U & 4400, 4401 \\
& & & 4400, 4400, 4399 \\ 
\hline
\vspace*{0.1cm}
OM & Image + Fast & B & 4400, 4400 \\
& & & 4401, 4399, 4399 \\ 
& & & 4400, 4400, 1260 \\ 
\hline
\end{tabular}
\end{center}
\vspace*{-0.3cm}
\caption{\label{tb:XMMObs} Details of the instruments, filters and uncorrected exposure times from the \textit{XMM-Newton} observation, 0935191701, considered here. The short Optical Monitor (OM) exposures occur while European Photon Imaging Camera (EPIC) is active, resulting in simultaneous optical (V and B band) and X-ray data. We note, however, that the U band exposures all occurred before the EPIC exposures.}
\end{table}

\begin{figure*}
    \centering
    \includegraphics[width=1.0\textwidth]{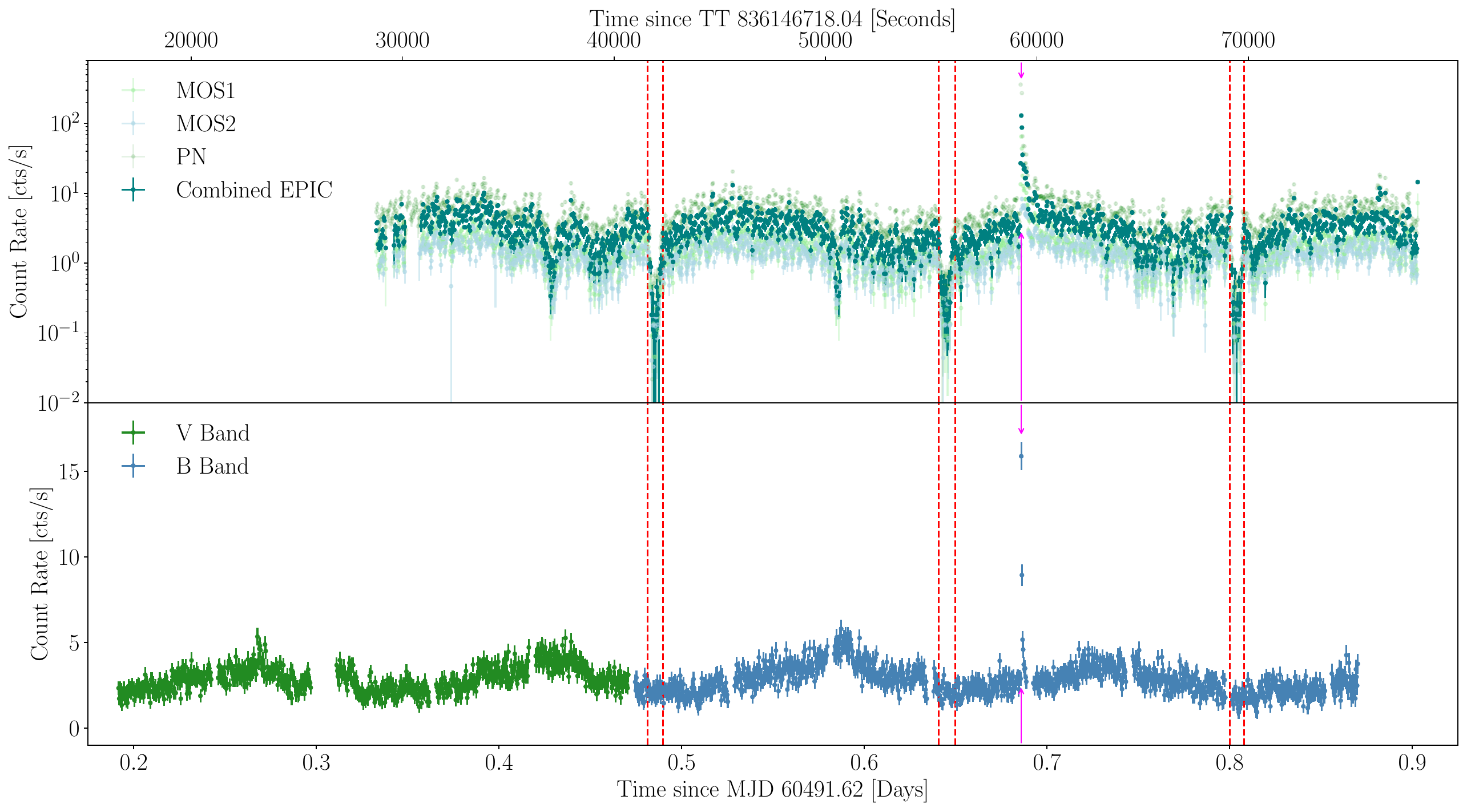}
    \vspace*{-0.25cm}
    \caption[]{\label{fig:full_lc} Top: $0.5 - 10$ keV combined EPIC light curve of ObsID 0935191701. The individual contributions from MOS1, MOS2 and PN are shown for comparison. Bottom: Full V and B band light curve from the Optical Monitor using the exposures that are coincident with the Combined EPIC light curve. In both panels, the pairs of dashed red lines mark the start of the X-ray ingress and end of the egress respectively and the magenta arrows mark the type I X-ray burst/optical burst peaks. Note the top and bottom x-axes are identically scaled.}
\end{figure*}

From \textit{XMM-Newton} ObsID 0935191701, we reduce all EPIC and OM exposures with the XMM-Newton Science Analysis Software, \textsc{xmm-sas v21.0.0} in conjunction with \textsc{heasoft v6.33} and the latest calibration files. We create EPIC-PN and EPIC-MOS event lists using the routines \texttt{epproc} and \texttt{emproc} respectively. Background flaring events are removed from these events lists by extracting high-energy light curves using \texttt{evselect} with the selection expression \texttt{\#XMMEA$\_$EP \&\& (PI>10000\&\&PI<12000) \&\& (PATTERN==0)} for EPIC-PN and \texttt{\#XMMEA$\_$EM \&\& (PI>10000) \&\& (PATTERN==0)} for EPIC-MOS. We determine good time intervals (GTIs) using \texttt{tabgtigen}, identifying a count rate threshold just above the background level. We subsequently apply barycentric corrections using \texttt{barycen} (to TDB using the default DE200 ephemeris) and extract $0.5 - 10.0$ keV source light curves with $1$s, $10$s and $30$s time bins. We use a $20$ arcsec circular region for MOS1, a $25$ arcsec circular region for MOS2 and a rectangular region of \texttt{(RAWX>=31) $\&\&$ (RAWX<=45)} for PN. Subsequently, background subtraction and corrections are performed using \texttt{epiclccorr}. The time axes of the corrected light curves are aligned and then added to obtain a combined EPIC light curve with the corresponding errors added in quadrature. 

The optical monitor (OM) observed simultaneously with EPIC, cycling through the U, V and B filters. However, we note that none of the U band exposures coincide with the EPIC exposures. For this reason we utilise only the V and B band data when producing the optical light curve. We reduce each V and B band OM exposure using the \textsc{xmm-sas} processing chain \texttt{omfchain}, leaving most parameters to their default values \footnote{\url{https://www.cosmos.esa.int/web/xmm-newton/sas-thread-omf}}. When extracting the light curves we apply circular source regions of 6 arcsec radius around the optical counterpart and use $1$s, $10$s and $30$s time bins to enable comparison to the combined EPIC X-ray light curve. As above, barycentric corrections are applied with \texttt{barycen}. 

The full X-ray and optical light curves are shown in Figure \ref{fig:full_lc}, where we see the X-ray eclipses marked with the red dashed lines. The type I X-ray burst, which peaks at $60492.309$ MJD, and corresponding optical burst are marked with magenta arrows. Both the optical and X-ray light curves display significant count rate modulations and appear anti-correlated; peaks in the X-ray modulation roughly coincide with troughs in the optical modulation. In addition, the X-ray light curve displays dips before the first two eclipses which arise due to the occultation of the X-ray emitting region by some optically thick medium e.g. accretion disc wind, hot spot or ablated material. These dips occur roughly $4500$s and $4700$s before their respective X-ray eclipses. There is no clear, single dip before the third X-ray eclipse. Instead the light curve displays a period of increased variability, during which there appear to be a series of smaller dips. The biggest dip occurs roughly $2700$ s before the third eclipse. During its previous outburst, EXO 0748--676 was found to display optical eclipses in phase with the X-ray eclipses \citep{Crampton1986}, but we do not identify the optical eclipses here. 

\section{X-ray and Optical Detection of the Type I Burst}
\label{sx:Burst}
\begin{figure}
    \centering
    \includegraphics[width=1.0\columnwidth]{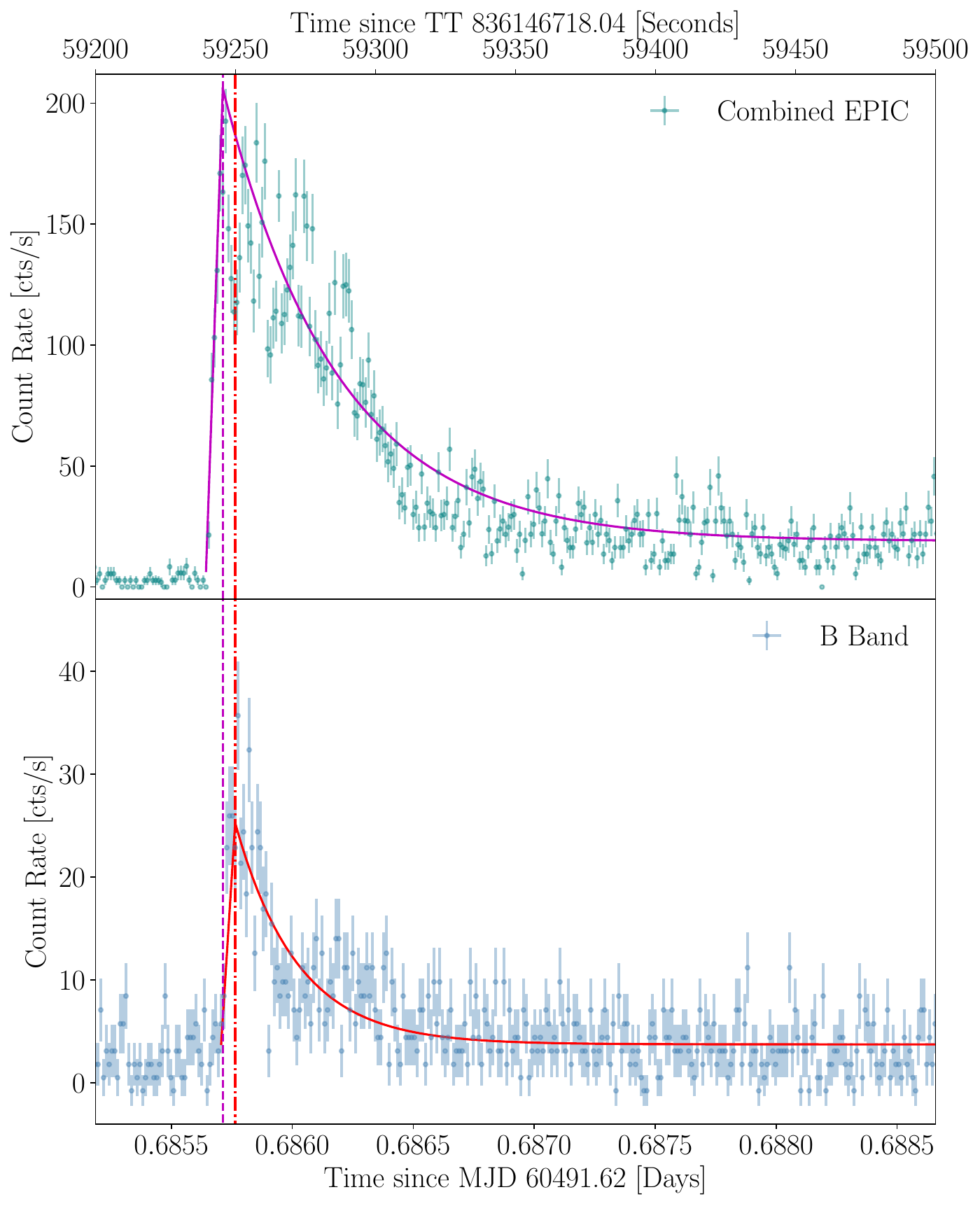}
    \vspace*{-0.25cm}
    \caption[]{\label{fig:Burst_lc} Fast rise -- exponential decay (FRED) fits to the type I X-ray burst as seen in the combined EPIC light curve (top) and the OM, B-filter light curve (bottom). The corresponding fit statistics are $\chi^{2} = 1.83$ and $\chi^{2} = 0.64$ respectively. The optical peak lags the X-ray peak by $4.46 \pm 1.71$s, as shown by the difference between the magenta dashed (X-ray peak) and red dot-dashed (optical peak) vertical lines. Note the top and bottom x-axes are identically scaled, but presented with different units.}
\end{figure}

As shown in Figure \ref{fig:full_lc}, the type I X-ray burst in the combined EPIC light curve \citep{Bhattachrarya2024Paper, Bhattacharya2024Atel} is coincident with a burst captured by the OM during an exposure with the B filter (exposure S018). Optical bursts have been observed to accompany X-ray bursts in only a handful of sources, including EXO 0748--676 (see e.g. \citealt{Paul2012}), and likely arise due to the reprocessing of the X-ray emission into the optical band \citep{Degenaar2018}. As such, there is typically a time delay between the original type I X-ray burst and the corresponding optical burst. To determine the time delay between the bands, we fit a simple FRED model consisting of a linear rise and an exponential decay, to the combined EPIC and OM burst profiles with $1$s time binning, and show the resulting fits in Figure \ref{fig:Burst_lc}. The linear rise is fit between the start of the burst and the peak, while the exponential decay component is fit to the first $250$s of data after the peak. Here, the start of the X-ray burst is defined as the first time before the peak when the count rate reaches $10$ per cent of the peak (e.g. \citealt{Albayati2023}). The same definition is used for the optical burst, except with a $20$ per cent threshold to account for the optical variability. From these fits we find that the peak of the optical burst lags that of the X-ray burst by $4.64 \pm 1.71$ s. This is consistent with the X-ray to optical burst lags measured by \citet{Paul2012}, who quote an average lag  of $3.25$s, although their measured lag times are as high as $8.0$s and the corresponding uncertainties are typically a few seconds. The start of the optical burst is also determined to lag the X-rays by $\approx 5$s. The modeled rise time from zero of the X-ray burst is $5.54 \pm 1.10$s while the modeled rise of the optical burst is slightly shorter at $4.13 \pm 1.09$s. As seen in Figure \ref{fig:Burst_lc}, the decay time of the X-ray burst is significantly longer than that of the optical burst returning to pre-burst levels after $\gtrsim 250$ s. In contrast, the optical burst returns to pre-burst rates after $\lesssim 100$s. To robustly compare the decay timescale between the two bands, we compute the e-folding time from the exponential decay portion of our FRED fits, finding $65 \pm 1$s and $48 \pm 6$s for the X-ray and optical e-folding time, respectively.

Our measured timescales are inconsistent with \citet{Paul2012}, who found both the rise times and the decay times of the bursts to be longer in the optical band. However, they also find that the burst profiles and timescales vary significantly; they measure optical rise times ranging from a few seconds to over $20$ seconds. We further note that the optical/X-ray burst presented here appears similar to that presented as event number 43 in \citet{Paul2012} (see their Figure 2), which has a weak optical component and decays faster than the corresponding X-ray burst. Thus, the difference in our measured decay times may be a result of a highly variable population of optical/X-ray burst events, for which the variability is likely related to the medium reprocessing the X-ray emission. Alternatively, the difference may be a consequence of observing a variable fraction of the X-ray luminosity within the \textit{XMM-Newton} bandpass -- the ratio of X-ray flux:counts will change with the temperature of the burst, thus affecting the profile of the X-ray burst compared to the flux driving the optical burst. Therefore, the difference in decay times is also explainable if the X-ray flux starts high and drops sharply. This can be tested by studying the energy-dependent light curves as the gas will respond differently to the soft and hard X-rays, although doing so is beyond the scope of this letter. A larger bandpass effect is typically seen in the optical/UV since the reprocessed peak moves from the UV to the optical as the burst cools \citep{Hynes2006} which is expected to sustain the optical burst, although this is contrary to our findings.


The component of the binary system reprocessing the X-ray burst emission into the optical band is likely located within a few light seconds of the X-ray emitting region \citep{Hynes2006, Paul2012}. Therefore, the accretion disc, the ablated outflow and the surface of the companion star are all possibilities but it is challenging to distinguish between them. The magnitude and variation of the time delays between the X-ray and optical emission informs on the location of the reprocessing site and thus the system's geometry. Broadly speaking, larger lags suggest a reprocessing site further from the X-ray emitting region (e.g. ablated material or companion star) while shorter lags favour a closer site (e.g. accretion disc or disc wind). In reality, however, lags from the disc or a disc wind will cover a broad range of timescales thus the magnitude of lags alone does not distinguish between reprocessing sites. If the properties of the bursts have some relation to the orbital phase of the system, this would suggest that the reprocessing site is phase-locked e.g. the irradiated face of the companion star. Alternatively, a relationship with luminosity would instead indicate that the reprocessing site is related to accretion e.g. a disc wind.

For the 63 X-ray/optical bursts detected by \textit{XMM-Newton} in EXO 0748--676's previous outburst, \citet{Paul2012} measure lags ranging from a few seconds to nearly ten seconds. However, the uncertainties associated with their measured lag times makes it impossible to distinguish between the possible reprocessing sites and the vast majority are consistent with the light crossing time of the orbit ($\approx 3$ s; \citealt{Hynes2006}). The same is true for the X-ray/optical burst presented here, as the measured lag time of $4.64 \pm 1.71$ s, is consistent with the light crossing time. Also, we cannot rule out that multiple independent sites are responsible for reprocessing the X-ray bursts, which could explain some of the variations in measured lags. 
Reprocessing could come from either Compton down-scattering \citep{Aftab2019} or absorption and re-emission \citep{Hynes2006}. We note, however, that for Compton down-scattering to produce optical from X-ray photons, multiple scatterings are required. In both cases, depending on the geometry of the reprocessor, the lag could be associated with light travel time. Additionally, if absorption and re-emission is responsible then an additional lag can come from the recombination time of the gas \citep{Silva2016} and this can dominate if the light travel time is relatively short (i.e. the reprocessor is close-in).
For our measured lag time of $4.46$ s, we can compute the required electron density using the Case B recombination coefficient for Hydrogen at $10,000$ K, $\alpha_{\rm{B}} = 3 \times 10 ^{-14}$. The timescale is given as $1 / (\alpha n_{e}$), thus the corresponding electron density is $1 \times 10 ^{13}$ cm$^{-3}$, which is easily achievable in a disc wind/atmosphere or corona.

Based on the results of \citet{Cominsky1987}, our measured lag time is consistent with a reprocessing site on the stellar surface. The authors find that the optical emission should appear $0.2$s plus the light crossing time after the initial X-ray emission, assuming that the deposition of X-ray radiation instantly heats the ionized stellar atmosphere. However, there is an existing argument against the irradiated face of the companion star being the reprocessing site based on the results of \citet{Paul2012}, who found no strong evidence of orbital phase dependence. However, the authors do not set an upper bound on the possible modulation or estimate the modulation expected from the companion star. Furthermore, only a few seconds of delay modulation is needed to be comparable to the light crossing time of the orbit ($\approx 3$ s; \citealt{Hynes2006}), so reprocessing on the surface of the companion is still possible. 

EXO 0748--676 was found to be undergoing irradiation driven ablation during its previous outburst, giving rise to material close to the companion star \citep{Knight2022a}. Thus, the ablated material could be the reprocessing site, although some orbital phase dependence would be expected as the material closest to the companion is the most optically thick. There is evidence that the ablated material moves and has previously reversed the asymmetry of the X-ray eclipses in EXO 0748--676 \citep{Knight2023}, so any phase dependence of the burst properties due to the ablated outflow may be more subtle. 

\begin{figure}
    \centering
    \includegraphics[width=\columnwidth]{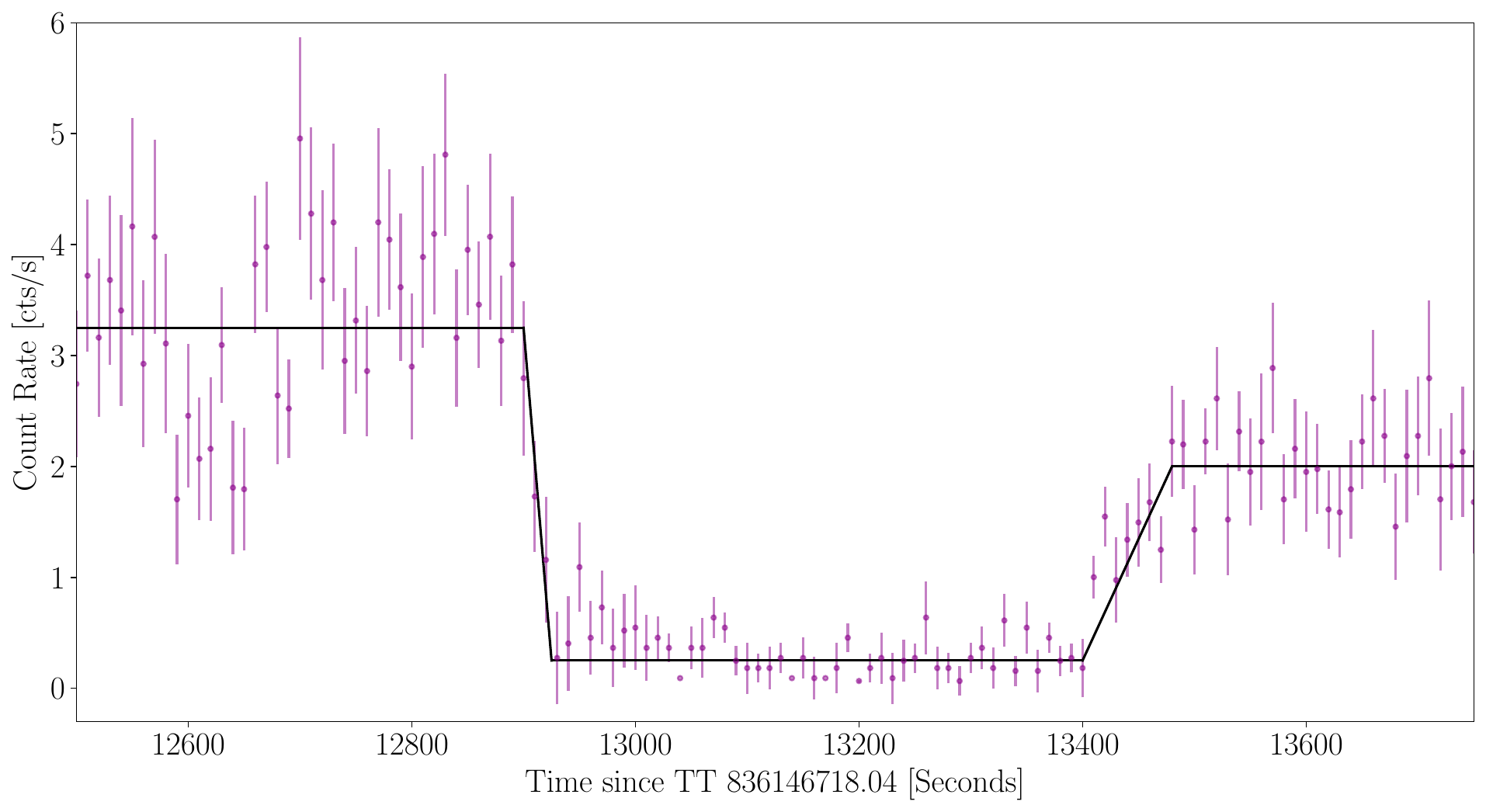}
    \vspace*{-0.5cm}
    \caption[]{\label{fig:eclipses} Combined EPIC phase-folded and averaged X-ray eclipse profile, fit with a simple eclipse model. The total duration of the eclipse is $\sim 585$s, consisting of a $25 \pm 5$s ingress, an $80 \pm 15$s egress and a $480 \pm 10$s totality.}
\end{figure}

To determine whether the ablated material is still present as EXO 0748--676 emerges from quiescence, and whether it could be the reprocessing site for the burst presented in Figures \ref{fig:full_lc} and \ref{fig:Burst_lc}, we measure the eclipse timings using our previously published simple eclipse model (\citealt{Knight2023}; see also \citealt{Wolff2009}). As EXO 0748--676 is a false widow, ablation should have continued throughout quiescence, being driven by a pulsar wind rather than X-ray irradiation \citep{Knight2023}. We expect the ablated material around the companion star to increase during quiescence since the pulsar wind mechanism is likely more efficient than the X-ray irradiation mechanism and a minimal amount of ablated material will be accreted during quiescence. \footnote{Spider pulsars are often found to host massive neutron stars \citep{Romani2012, Linares2018, Burdge2022}, which may be due to the accretion of ablated material.}. Some of the material may be driven outwards, perhaps forming a circumbinary disc \citep{Antoniadis2014}, but the pulsar wind will continually liberate material from the stellar surface. As such the X-ray eclipses should still be asymmetric and extended in time, perhaps even more so than when it finished its first outburst in 2008. The solid line in Figure \ref{fig:eclipses} shows the fit to the average, phase-folded eclipse profile with $10$s time bins. Here the later two X-ray eclipses are folded onto the time of the first X-ray eclipse. The total duration of the eclipse is $\sim 585$s, consisting of a $25 \pm 5$s ingress, an $80 \pm 15$s egress and a $480 \pm 10$s totality. Overall, the eclipses are consistent with those seen in the first outburst, occurring once per orbital period. However, the eclipse transitions are significantly more extended than before (see \citealt{Knight2022a}), supporting the hypothesis that EXO 0748--676 is a false widow \citep{Knight2022a, Knight2023} and was undergoing pulsar wind-driven ablation while in quiescence. This also means that the ablated material is a possible reprocessing site for the optical/X-ray burst presented here. While indicative of ongoing ablation, we note that the simple eclipse model is unable to fully capture the shape of the eclipse profile or account for variability from the primary source. The eclipse profiles exhibited by EXO 0748--676 are known to be much more complex (see e.g. \citealt{Knight2022a}), thus, a detailed analysis of the eclipses is required for more precise measurements, but is beyond the scope of this letter.

\vspace{-0.5cm}
\section{Conclusions}
\label{sx:Conclusion}
We have presented the first simultaneous optical and X-ray detection of a type I X-ray burst during the new outburst of EXO 0748--676. The type I X-ray burst, captured by the EPIC detectors, occurred at $60492.309$ MJD. The OM captured a corresponding peak $4.46 \pm 1.71$s later, during an exposure with the B filter in image $+$ fast mode, and is consistent with a scenario whereby the X-ray burst emission is reprocessed into the optical band \citep{Degenaar2018, Paul2012}. We determine that the X-ray and optical bursts have similar rise times, $\approx 5.5$s and $\approx 4.1$s, respectively, but different decay timescales. The e-folding times are found to be approximately $65$s and $48$s for the X-ray and optical, respectively.
While it is typical for the optical burst decay to be longer in duration than the X-ray burst decay, shorter optical decays have been observed when the optical component is weak \citep{Paul2012}, similar to that found here. We note that observing the optical emission to be lagged but not smeared out in anyway is unusual and will require further exploration. As determined previously, the reprocessing site likely resides within a few light seconds of the X-ray emitting region. Thus, the companion star, accretion disc and ablated outflow are all possible. Distinguishing between the possible reprocessing sites confidently will require further observations of simultaneous optical and X-ray bursts, which in turn, will allow us to assess whether dynamic or geometric effects influence the properties of the reprocessed bursts. Furthermore, multi-colour data will be valuable when determining the temperature and size of the reprocessing site, while spectral coverage could tell us if we are seeing optically thin recombination from ablated material. Such information is crucial to understanding the origin of reprocessed bursts.

\vspace*{-0.5cm}
\section*{Acknowledgements}
The authors acknowledge and thank the \textit{XMM-Newton} team and Dr. Norbert Schartel for promptly observing the target through Director's Discretionary Time.

TPR and AHK acknowledge support from the Science and Technology Facilities Council (STFC) as part of the consolidated grant award ST/X001075/1. AI and JHM acknowledge support from the Royal Society. MM acknowledges support from the Science and Technology Facilities Council (STFC) as part of the consolidated grant award ST/V001000/1. NCS acknowledges support from the Science and Technology Facilities Council (STFC) grant ST/X001121/1.

\vspace*{-0.5cm}
\section*{Data Availability}
The data used in this study are publicly available via the \textit{XMM-Newton} Science Archive. All software used in this study is freely available online. 



\vspace*{-0.5cm}
\bibliographystyle{mnras}
\bibliography{All_Refs} 


\bsp 
\label{lastpage}
\end{document}